%% file: bad.tex
\input  macros.tex

\nopageonenumber
\baselineskip = 18pt
\barsoff



\def\vareps{\varepsilon}

\def\bk{\item{}}

\def\eg{{\it e.g.,}\ }
\def\ie{{\it i.e.,}\ }


\line{{\tt hep-th/9907188} \hfil McGill-98/35}

\title
\centerline{The Free Energy of N=4 Super-Yang-Mills}
\vskip 0.1in
\centerline{and the AdS/CFT Correspondence}
\endtitle

\authors
\centerline{C.P. Burgess, N.R. Constable and R.C. Myers}  
\vskip .07in
\centerline{\it Physics Department, McGill University}
\centerline{\it 3600 University St., Montr\'eal, Qu\'ebec, 
CANADA, H3A 2T8.}
\endauthors

\vskip .25in

\abstract
We compute the high-temperature limit of the
free energy for four-dimensional $N=4$ supersymmetric
$SU(N_c)$ Yang-Mills theory. At weak
coupling we do so for a general ultrastatic background
spacetime, and in the presence of slowly-varying background
gauge fields. Using Maldacena's conjectured duality, we calculate the 
strong-coupling large-$N_c$ expression for the special case that the
three-space has constant curvature.
We compare the two results paying particular attention to curvature
corrections to the leading order expressions.
\endabstract


\ref\juanA{J. Maldacena, Adv. Theor. Math. Phys. {\bf 2} (1998) 231
[hep-th/9711200].}

\ref\igorA{S.S. Gubser, I.R. Klebanov and A.M. Polyakov,
Phys. Lett. {\bf B428} (1998) 105 [hep-th/9802109].}

\ref\wittenA{
E. Witten, Adv. Theor. Math. Phys. {\bf 2} (1998) 253 [hep-th/9802150].}

\ref\revue{
O.~Aharony, S.S.~Gubser, J.~Maldacena, H.~Ooguri and Y.~Oz,
``Large N field theories, string theory and gravity,''
hep-th/9905111.}

\ref\therm{
E. Witten, Adv. Theor. Math. Phys. {\bf 2} (1998) 505 [hep-th/9803131].}

\ref\christensen{S. Christensen, Ph.D. thesis, University of Texas, 1977.}

\ref\igorB{
S.S. Gubser, I.R. Klebanov and A.W. Peet, Phys. Rev. {\bf D54} (1996) 3915
[hep-th/9602135].}

\ref\arkady{
A.A.~Tseytlin and S.~Yankielowicz,
Nucl. Phys. {\bf B541} (1999) 145 [hep-th/9809032].}

\ref\moretherm{
N. Itzhaki, J.M. Maldacena, J. Sonnenschein, S. Yankielowicz,
Phys. Rev. {\bf D58} (1998) 046004 [hep-th/9802042];\bk
S-J. Rey, S. Theisen and J-T. Yee, Nucl. Phys. {\bf B527} (1998) 171
[hep-th/9803135];\bk
A. Brandhuber, N. Itzhaki, J. Sonnenschein and S. Yankielowicz,
JHEP {\bf 06} (1998) 001 [hep-th/9803263];\bk 
H. Dorn and H.J. Otto, JHEP {\bf 09} (1998) 021 [hep-th/9807093];\bk
A.~Chamblin, R.~Emparan, C.V.~Johnson and R.C.~Myers,
Phys. Rev. {\bf D59} (1999) 064010 [hep-th/9808177];\bk
S.W. Hawking, C.J. Hunter and D.N. Page, Phys. Rev. {\bf D59} (1999) 044033
[hep-th/9809035];\bk
J. L. F. Barbon, I. I. Kogan and E. Rabinovici, 
Nucl. Phys. {\bf B544} (1999) 104 [hep-th/9809033];\bk
A. Hashimoto and Y. Oz, Nucl. Phys. {\bf B548} (1999) 167
[hep-th/9809106];\bk
S. Kalyana Rama and B. Sathiapalan, 
Mod. Phys. Lett. {\bf A13} (1998) 3137 [hep-th/9810069];
[hep-th/9905219];\bk
A.W. Peet and S.F. Ross, JHEP {\bf 12} (1998) 020 [hep-th/9810200];\bk
Y. Kinar, E. Schreiber and J. Sonnenschein,  
[hep-th/9811192];\bk 
G. Ferretti and D. Martelli, [hep-th/9811208];\bk
S.S. Gubser, Nucl. Phys. {\bf B551} (1999) 667
[hep-th/9810225];\bk
R.-G. Cai and K.-S. Soh, [hep-th/9812121];
JHEP {\bf 05} (1999) 025 [hep-th/9903023];\bk
J. Greensite and P. Olesen, JHEP {\bf 04} (1999) 001 [hep-th/9901057];\bk
K.~Landsteiner, Mod. Phys. Lett. {\bf A14} (1999) 379 [hep-th/9901143];\bk
A.~Chamblin, R.~Emparan, C.V.~Johnson and R.C.~Myers, to appear
in Physical Review D15, [hep-th/9902170]; [hep-th/9904197];\bk
J.~Ellis, A.~Ghosh and N.E.~Mavromatos,
Phys. Lett. {\bf B454} (1999) 193 [hep-th/9902190];\bk
M.~Cvetic and S.S.~Gubser,
JHEP {\bf 04} (1999) 024 [hep-th/9902195]; JHEP {\bf 07} (1999) 010
[hep-th/9903132];\bk
M.M.~Caldarelli and D.~Klemm, [hep-th/9903078];\bk
A.~Brandhuber and K.~Sfetsos, [hep-th/9906201];\bk
S.~Nojiri and S.D.~Odintsov, [hep-th/9906216];\bk
E.~Kiritsis and T.R.~Taylor, [hep-th/9906048];\bk
E.~Kiritsis, [hep-th/9906206];\bk
D.S.~Berman and M.K.~Parikh, [hep-th/9907003].}

\ref\igorC{
S.S. Gubser, I.R. Klebanov and A.A. Tseytlin, Nucl. Phys. {\bf B534}
(1998) 202 [hep-th/9805156].}

\ref\taylor{
A. Fotopoulos and T.R. Taylor, Phys. Rev. {\bf D59} (1999) 061701
[hep-th/9811224];\bk
M.A.~Vazquez-Mozo, ``A Note on supersymmetric Yang-Mills thermodynamics''
[hep-th/9905030];\bk
C.~Kim and S.~Rey, ``Thermodynamics of large N super-Yang-Mills theory
and AdS/CFT correspondence'' [hep-th/9905205];\bk
A.Nieto and M.Tytgat,[hep-th/9906147].}

\ref\phase{
M. Li, JHEP {\bf 03} (1999) 004 [hep-th/9807196];\bk
Y. Gao and M. Li, Nucl. Phys. {\bf B551} (1999) 229 [hep-th/9810053].}

\ref\WbgGravity{
S. Weinberg, {\it Gravitation and Cosmology}, Wiley (1972).}

\ref\garyA{
G.T. Horowitz and R.C. Myers, Phys. Rev. {\bf D59} (1999)
026005 [hep-th/9808079].}

\ref\basic{F. Gliozzi, J. Scherk and D. Olive Nucl. Phys. B122 (1977)
253. For a review see: M.F. Sohnius, Phys. Rep. 128 (1985) 2.}
 
\ref\scale{
Enrico C. Poggio and H.N. Pendleton, 
Phys. Lett. 72B (1977) 200;\bk
Marc Grisaru, Martin Rocek and Warren Siegel,
Phys. Rev. Lett. 45 (1980) 1063-1066; 
Nucl. Phys. B183 (1981) 141;\bk
William E. Caswell and Daniela Zanon,
Nucl. Phys. B182 (1981) 125;\bk
Martin F. Sohnius and Peter C. West, 
Phys. Lett. 100B (1981) 245;\bk
K.S. Stelle and P.K. Townsend, Phys. Lett. 113B (1982) 25;\bk
L.V. Avdeev and O.V. Tarasov, 
Phys. Lett. 112B (1982) 356-358.}

\ref\gilk{P.B. Gilkey, J. Diff. Geom. {\bf 10} (1975) 601;
Adv. Math. {\bf 15} (1975) 334.}

\ref\physrep{
For a review see: 
A.O. Barvinsky and G.A. Vilkovisky, Phys. Rep.
{\bf 119} (1985) 1.}

\ref\Schwinger{J.S. Schwinger, Phys. Rev. {\bf 82} (1951) 664}

\ref\DTGF{
B.S. DeWitt, `The dynamical theory of groups and fields' in {\it Relativity,
Groups and Topology}, eds. B.S. DeWitt and C. DeWitt (New York, Gordon and
Breach, 1965).}

\ref\dowker{
J.S. Dowker and G. Kennedy, J. Phys. A: Math. Gen. 11 (1978) 895.}

\ref\old{
E.S.~Fradkin and A.A.~Tseytlin, Phys. Lett. {\bf 134B} (1984) 187.}

\ref\wandw{
E.T. Whittaker and G.N. Watson, {\it A Course of Modern Analysis},
Cambridge University Press.}

\ref\WbgFT{S. Weinberg, in the proceedings of the International
School of Subnuclear Physics, Erice, 1976, A. Zichichi ed., Plenum Press.}

\ref\FTeffact{See, for example, 
J. Kapusta, {\it Finite Temperature Field
Theory}, Cambridge University Press, 1989;\bk 
M. Quiros, {\it Helv. Phys. Acta.} {\bf 67} (1994) 452--583,
and references therein.}

\ref\mtw{
C.W. Misner, K.S. Thorne and J.A. Wheeler, {\it Gravitation} (San Francisco,
Freeman, 1973).}

\ref\yuri{
Related infrared singularities in curved spaces are discussed
by Y.V. Gusev and A.I. Zelnikov, ``Finite temperature nonlocal effective
action for quantum fields in curved space,'' [hep-th/9807038].}

\ref\birmbh{D.~Birmingham,
Class. Quant. Grav. {\bf 16} (1999) 1197 [hep-th/9808032].}

\ref\don{
S.W. Hawking and D.N. Page, Commun. Math. Phys. {\bf 87} (1983) 577.}

\ref\garyB{
S.W. Hawking and  G.T. Horowitz, Class. Quant. Grav. {\bf 13}
(1996) 1487 [gr-qc/9501014].}

\ref\roberto{
R.~Emparan, JHEP {\bf 06} (1999) 036 [hep-th/9906040].}

\ref\gluondamping
{U. Heinz, K. Kajantie and T. Toimela, {\it Phys. Lett.}
{\bf 183B} 96 (1987);\bk
R. Kobes and G. Kunstatter, {\it Physica} {\bf  A158} 192-199 (1989);\bk 
Eric Braaten and Robert D. Pisarski, {\it Phys. Rev. Lett.}
{\bf 64} 1338 (1990); {\it Phys. Rev.} {\bf D42} 2156-2160 (1990);\bk 
V.V. Lebedev and A.V. Smilga, {\it Phys. Lett.} {\bf B253} 231-236
(1991);\bk
C.P. Burgess and A.L. Marini, {\it Phys. Rev.} {\bf D45}
17-20 (1992) [hep-th/9109051];\bk
Anton Rebhan, {\it Phys. Rev.} {\bf D46} 482-483 (1992) 
[hep-ph/9203211];\bk 
F. Flechsig, A.K. Rebhan and H. Schulz, {\it Phys. Rev.}
{\bf D52} 2994-3002 (1995) [hep-ph/9502324];\bk 
Xiao-Fei Zhang and Jia-Rong Li, {\it Phys. Rev.} {\bf C52}
964-974 (1995); {\it Commun. Theor. Phys.} {\bf 26}
339-346 (1996);\bk
De-Fu Hou and Jia-Rong Li, {\it Sci. Sin.} {\bf A39} 540-546 (1996);\bk 
Ziao-ping Zheng and Jia-rong Li, {\it Phys. Lett.} {\bf B409}
45-50 (1997) [hep-th/9707052];\bk
De-fu Hou and Jia-rong Li, {\it Nucl. Phys.} {\bf A618}
371-380 (1997) [nucl-th/9707033];\bk
A. Abada, O. Azi and K. Benchallal , {\it Phys. Lett.}
{\bf B425} 158-165 (1998) [hep-ph/9712210];\bk
Ji-sheng Chen and Jia-rong Li, {\it Phys. Lett.} {\bf B430}
209-216 (1998) [hep-ph/9803455];\bk
By Yuri A. Markov and Rita A. Markova, ``The Problem of Nonlinear 
Landau Damping in Quark Gluon Plasma'' preprint [hep-ph/9902397];\bk
M. Dirks, A. Niegawa and K. Okano, ``A Slavnov-Taylor Identity and
Equality of Damping Rates for Static Transverse and Longitudinal Gluons
in Hot QCD'', preprint [hep-ph/9907439].}

\ref\GBIR
{R.Pisarski and M.Tytgat, {\it Phys. Rev. Lett.} {\bf 78} 3622-3625 (1997) 
[hep-ph/9611206];\bk
C.Manuel, {\it Phys. Rev.} {\bf D57} 2871-2878 (1998) [hep-ph/9710208];\bk
G. Alexanian, E.F. Moreno, V.P. Nair and R. Ray,
{\it Phys. Rev.} {\bf D60} (1999) 011701 [hep-ph/9903232].}

\ref\GBreview{For a recent review of goldstone boson properties
see C.P.~Burgess, ``A Goldstone Boson Primer'',
to appear in Physics Reports [hep-ph/9812468 ].}

\ref\Dorey{
N. Dorey, ``An elliptic superpotential for softly broken $N=4$ supersymmetric
Yang-Mills theory'', preprint UW/PT 99-10 [hep-th/9906011].}

\ref\boundterms{
M.~Henningson and K.~Skenderis, JHEP {\bf 07} (1998) 023
[hep-th/9806087];\bk
V.~Balasubramanian and P.~Kraus,
``A Stress tensor for Anti-de Sitter gravity''
[hep-th/9902121];\bk
P.~Kraus, F.~Larsen and R.~Siebelink,
``The gravitational action in asymptotically AdS and flat space-times,''
hep-th/9906127.}

\ref\boundtermb{
Roberto Emparan, Clifford V. Johnson and Robert C. Myers, 
``Surface terms as counterterms in the AdS/CFT correspondence,''
to appear in Physical Review D15 [hep-th/9903238].}

\ref\boundman{
R.B.~Mann, ``Misner string entropy'' [hep-th/9903229];\bk
S.R.~Lau, ``Light cone reference for total gravitational energy''
[gr-qc/9903038].}

\ref\uvir{
L.~Susskind and E.~Witten, ``The Holographic bound in anti-de Sitter space''
[hep-th/9805114].}



\vfill\eject
\section{Introduction}

There is substantial evidence to support Maldacena's conjectured
duality relating superstring theory in anti-de Sitter (AdS) spacetime
and a conformal field theory (CFT) \juanA\igorA\wittenA
--- for a comprehensive review, see ref.~\revue. In particular,
type IIB superstring theory on $AdS_5\times S^5$ is dual to 
four-dimensional $N=4$
super-Yang-Mills (SYM) theory with gauge group $SU(N_c)$ \juanA.
An interesting aspect of this duality is the study of the behavior of
the SYM theory at finite temperature
\juanA\wittenA\therm\igorB\arkady\moretherm\igorC\taylor\phase.
In the superstring theory
for sufficiently high temperatures, the thermal state is described
by an asymptotically AdS black hole \juanA\wittenA. 
One finds that, at a qualitative level, one recovers many of the expected
physical properties of the thermal SYM system from the black hole geometry.
A quantitative comparison is more difficult because supergravity
provides a description of the SYM theory at strong coupling, and so
no direct calculations can be made reliably in the SYM theory. In some
cases \igorB\garyA, \eg\ the free energy density, perturbative calculations at
weak coupling still reproduce
the strong coupling results up to factors of order one. So it would
appear that the corresponding coefficients are smooth functions of the
effective coupling which interpolate between the strong and weak coupling
results. In the case of the free energy density, subleading
corrections have been calculated at both strong \igorC\ and
weak \taylor\ coupling, and the results are suggestive that the interpolation
may even be achieved by simply a monotonic function.  

Whether or not there is a smooth interpolation between
strong and weak coupling is a question deserving close scrutiny.
By carefully examining the expansions of the free energy at small
and large coupling, and by making some simple assumptions about the
behavior of the expansion coefficients, Li \phase\ has argued that it is
impossible to smoothly interpolate between these two coupling regimes
of the SYM theory. Hence Li concluded that there must be a phase
transition at some critical coupling when the SYM theory
is at any finite temperature. The primary motivation for the present
paper was to investigate the free energy calculations in more detail
to look for evidence of such a phase transition. As well
as introducing a finite temperature, we consider the SYM theory on a
curved background space. Finite temperature is included
by working with a four-dimensional, ultrastatic Euclidean spacetime 
for which the Euclidean `time' direction is periodically identified
with period $\beta = 1/T$. For the spatial geometry,
we pay particular attention to three-geometries of constant curvature
$\kappa/\ell^2$ (with $\kappa = 0,\pm 1$)
since these are the cases for which we may also compute the strong-coupling
free energy using the AdS/CFT correspondence. 
When comparing the results obtained for weak and strong couplings, 
we follow how they depend on the radii, $\beta$ and $\ell$, in
the limit $\beta/\ell \ll 1$ (which corresponds 
to the high-temperature limit). 

We organize the paper as follows: Section 2 states the
preliminaries, describing the $N=4$ super-Yang-Mills theory.
Section 3 then addresses the weak-field
calculation, which may be performed quite generally, using 
well-known heat-kernel techniques.  
We compute the weak coupling effective action for SYM at finite temperature,
and in the 
presence of slowly-varying background gauge and gravitational fields.
The strong-coupling string/supergravity 
calculation is then given in section 4. A brief discussion of
our results is given in section 5.

\section{$N=4$ Super-Yang-Mills Theory}

The field content of $N=4$ SYM theory \basic\ is
$\{ A^a_\mu, \lambda^a_i, \varphi^a_r \}$, where $a=1,\dots,
d_\ssg$ runs over the adjoint representation of the gauge group,  $G$,
$i=1,\dots,4$ counts the spin-half fields, and $r = 1,\dots,6$ 
labels the scalars. Where necessary we will choose $G=SU(N_c)$,
for which $d_\ssg = N_c^2 - 1$, although this choice does not play
an important role in the weak coupling calculations. 

The action of the theory may be conveniently formulated in terms
of $N=1$ superfields, of which we require one gauge multiplet,
$\Scw_\ssl^a$, and three matter multiplets, $\phi^a_m$, all in the
adjoint representation. The action is then given by minimal
kinetic terms for all fields, gauge interactions, plus those
interactions derived from the $N=1$ superpotential:
\label\NfourW
\eq
W = {i \sqrt2 \over 3!} \; \epsilon^{mnp} \; c_{abc} \;
\phi^a_m \, \phi^b_n \, \phi^c_p. 
\eeq
Here $c_{abc}$ represent the completely antisymmetric structure
constants for the gauge group. 

This superpotential is manifestly
invariant under a global $SU(3)$ flavour symmetry (acting on the 
indices $m, n$ and $p$), as well as a $U(1)$
$R$-symmetry for which the charge assignments of the fields are
$R(\Scw^a_\ssl) = 1$ and $R(\phi^a_r) = {2\over 3}$. These are
subgroups of a larger $SU(4)$ flavour symmetry which
this model enjoys, defined as the automorphism of the supersymmetry
algebra which rotates the four supersymmetry generators amongst
themselves. Only the $SU(3) \times U_\ssr(1)$ subgroup is
manifest when the theory is expressed in terms of $N=1$
superfields. 

The theory is also scale invariant, even at the quantum level \scale. This
may be argued within perturbation theory using the nonrenormalization
theorems of $N=2$ supersymmetry, or by constructing the most
general $N=1$ supersymmetric action with this field content
which admits the above-mentioned $SU(4)$ symmetry. 

The theory's scalar potential is 
\label\scapot
\eq
V = \sum_{am} \left| {\partial W \over \partial \phi^a_m} \right|^2,
\eeq
whose minima are described by the supersymmetry-preserving
flat directions for which $c_{abc} \; \phi^b_n \, \phi^c_p = 0$.
Semiclassically, nonzero fields along these flat directions cost no
energy and spontaneously break
the gauge group, the global $SU(4)$ symmetry and scale invariance. A number
of Goldstone and massless gauge multiplets therefore figure prominently among
the low-energy states, at least at weak coupling.

In the following, we will examine the free energy of this theory at
finite temperature and in curved background spacetimes. 
A nontrivial result arises because both the temperature 
and background curvatures typically break supersymmetry, thereby
permitting a nonvanishing free energy. 
In all ways but one, the action we use for curved spacetimes
is the same as the one just described, albeit with the substitution
everywhere of covariant (with respect to diffeomorphisms) derivatives.
The only nonminimal change
required is that the scalars couple conformally to the Ricci scalar
\wittenA.
This ensures the conformal invariance of the theory for general
background metrics.

\vfill\eject
\section{Weak-Coupling Calculation}

We now outline the weak-coupling calculation of the effective action
and free energy density. Our approach is to compute the contribution
of the short-distance, ultraviolet part of the theory to the free energy
using the well-established 
Gilkey-DeWitt heat-kernel methods \Schwinger \gilk \DTGF \christensen 
(see also \physrep), which we will briefly review. After the heat
kernel discussion we examine what inferences may be drawn from these
methods about the long-distance contributions to the free energy.

\subsection{Heat-Kernel Methods}

Heat-Kernel techniques are based upon the following representation of 
the functional determinants which appear at one loop. Assuming
Euclidean signature in an $n$-dimensional 
spacetime \Schwinger \DTGF \christensen,
\label\HKrep
\eq
\eqalign{
\Sigma &= \pm \, \hf \Tr \Log \Bigl( - \Square + m^2 + X \Bigr) \cr
&= \mp \, \hf \int dV \int_0^\infty { ds\over s} \; \tr \;
K(x,x;s) ,\cr}
\eeq
where the upper (lower) sign is for bosons (fermions),
$dV$ is the covariant 
volume element, and $K(x,y;s)$ satisfies the equation:
\label\heateq
\eq
{\partial K \over \partial s} + \Bigl( - \Square + m^2 + X
\Bigr) \; K = 0, 
\eeq
with initial condition $K(x,y;s) \to \delta^n(x,y)$ as $s \to 0$. 

Part of the utility of this representation lies in the observation
that the contribution of the short-distance part of the 
system to $\Sigma$ is controlled by the small-$s$ 
behaviour of $K(x,y;s)$. Furthermore,
this small-$s$ behaviour has been computed once and for all, for
general choices  for the operator $-\Square + X$. 

Concretely, if $K(x,y;s)$ is expanded for small $s$:
\label\akdefs
\eq
K(x,y;s) = K_0(x,y;s) \;\sum_{k = 0}^\infty a_k(x,y) \; s^k , 
\eeq
with $K_0(x,y;s)$ an appropriately-chosen function \christensen, then $\Sigma$
takes the following form:
\label\genSigform
\eq
\Sigma = \pm \; {1 \over 2 \, (4 \pi)^{n/2}} \; 
\sum_{k=0}^\infty c_k \;\int dV \;  \tr \; a_k(x,x) , 
\eeq
where $c_k$ represent the following integrals:
\label\ztints
\eq
\eqalign{
c_k &= \int_0^\infty {ds \over s} \;\; s^{k - n/2} \; \exp(- m^2 s)  \cr
&= m^{n-2k}\, \Gamma\left( k - {n \over 2} \right), \cr}
\eeq
and the first few $a_k(x,x)$ are given explicitly by 
\gilk \christensen:\foot\wbgc{In this section, we adopt the curvature
conventions of ref.~\WbgGravity.}
\label\firstfewak
\eq
\eqalign{
a_0(x,x) & = I, \cr
a_1(x,x) &= - \; \nth6 \; \Bigl(R + 6 X \Bigr), \cr
a_2(x,x) &= \nth{360} \; \Bigl( 2 \, R_{\mu\nu\lambda\rho}
R^{\mu\nu\lambda\rho} - 2 R_{\mu\nu} R^{\mu\nu} + 5 R^2 - 
12 \; \Square \, R \Bigr) \; I \cr
& \qquad\qquad + \nth6 \; R \; X + \hf \; X^2 - \nth6 \; \Square \, X
+ \nth{12} \; Y_{\mu\nu} Y^{\mu\nu}. \cr}
\eeq

In this last expression $Y_{\mu\nu}$ is defined by $Y_{\mu\nu} = 
\Bigl[ \nabla_\mu , \nabla_\nu \Bigr]$, where $\nabla_\mu$ is the
gauge and coordinate covariant derivative appearing in
$\Square = g^{\mu\nu}\nabla_\mu \nabla_\nu$.
In the presence of background gravitational
and gauge fields, this commutator becomes:
\label\Ymndef
\eq
Y_{\mu\nu} = -i F^a_{\mu\nu} \; t_a - {i \over 2} \; 
{R_{\mu\nu}}^{\alpha\beta}
\; J_{\alpha\beta},
\eeq
where $t_a$ and $J_{\alpha\beta}$ represent the generators of 
gauge and Lorentz transformations
on the field of interest. 

\subsection{Applications to Spins Zero, Half and One}

We next record the above expressions for the three spins of
interest in the present problem. For the moment we leave the dimension of
spacetime arbitrary, although --- with dimensional regularization in
mind --- we imagine evaluating $n = 4 - 2\vareps$ at the end of the
calculation. 
Our results are given in the presence of background metric and gauge
fields. (Background scalars are also implicitly included through the
dependence on the particle mass, $m$.)

\topic{Spin Zero} For spinless particles we consider the operator 
$-\Square + m^2_0 + \xi\, R$ acting on real scalar fields, 
where the choices $\xi = 0$ and $\xi = - \nth6$ respectively 
correspond  (in four dimensions) to a minimally-coupled 
and conformally-coupled scalar.  Denoting
the gauge generators as represented on scalars by $t_a$, and assuming
that we have $N_0$ scalars sharing the  same value of $\xi$ we find:
\label\Ymnspinzero
\eq
Y_{\mu\nu} = -i \; F^a_{\mu\nu} \; t_a ,
\eeq
and so the $a_k(x,x)$ become:
\label\akspinzeroa
\eq
\tr{}_0\; a_0(x,x)  = N_0,  \qquad  
\tr{}_0 \; a_1(x,x)  = - N_0 \; \left( \xi + \nth6 \right) \; R ,
\eeq
\label\akspinzerob
\eq
\eqalign{
\tr{}_0\; a_2(x,x)  &= N_0 \; \left[ \nth{180} \; R_{\mu\nu\lambda\rho}
R^{\mu\nu\lambda\rho} - \nth{180} \; R_{\mu\nu} R^{\mu\nu} +
\hf \; \left(\xi + \nth6 \right)^2  R^2 \right. \cr
& \qquad\qquad \qquad\qquad \left. - \; \nth6 \; \left(\xi + \nth5 \right)
\; \Square \, R \right] - \; \nth{12} \; C(\Scr_0) \; F^a_{\mu\nu}
F_a^{\mu\nu}.\cr}
\eeq
Here, and in what follows, $\Scr_s$ denotes 
the gauge representation of the particles of spin $s$, and the Dynkin
index, $C(\Scr)$, is defined by $\tr(t_a t_b) = C(\Scr) \; \delta_{ab}$.
We normalize the gauge generators so that $C(F) = \hf$ for the fundamental
representation of $SU(N_c)$. 
 
\topic{Spin One Half}
Without loss of generality we represent spin-half particles using
Majorana spinors. For such fields the operator of interest is $\delsl + m_\hf$,
for which 
the determinant is not well-defined. In the absence of anomalies in
the Lorentz or gauge group, we instead define this determinant 
as the square root
of the determinant of $\Bigl(- \delsl + m_\hf\Bigr)
\Bigl( \delsl + m_\hf \Bigr)$, implying
\label\sqrttrick
\eq
\eqalign{
\Tr \Log \Bigl(\delsl + m_\hf \Bigr) &= \hf \; \Tr \Log \Bigl( -\delsl^2 +
m_\hf^2\Bigr) \cr
&= \hf \Tr \Log \left[ - \Square + m_\hf^2
- \nth4 \; R + \frac{i}{2} \; 
\gamma^{\mu\nu} F^a_{\mu\nu} \, \left(T_a \gamma_\ssl - 
T^*_a \gamma_\ssr \right) \right]. \cr}
\eeq
Here $\gamma_{\mu\nu} = \hf \left[ \gamma_\mu, \gamma_\nu \right]$
and $T_a$ is the gauge generator as represented on left-handed fermions.

The previous formalism may now be applied to this expression, provided
the extra factor of `$\,\hf\,$' seen in eq.~\sqrttrick\ is kept in mind. The
traced commutator in this case is: 
\label\Ymnspinhalf
\eq
\tr \; \left( Y_{\mu\nu} Y^{\mu\nu} \right) = - \; \hf \; N_\hf
\; R_{\mu\nu\lambda\rho}
R^{\mu\nu\lambda\rho} -4 C\left( \Scr_\hf \right)  F^a_{\mu\nu} F_a^{\mu\nu},
\eeq
and so the $a_k(x,x)$ become:
\label\akspinhalfa
\eq
\tr{}_\hf\; a_0(x,x) = 2 N_\hf , \qquad 
\tr{}_\hf \; a_1(x,x) = \nth6 \; N_\hf  \; R ,
\eeq
\label\akspinhalfb
\eq
\eqalign{
\tr{}_\hf\; a_2(x,x) &= N_\hf \; \left[ -\; \frac{7}{720} \;
R_{\mu\nu\lambda\rho}
R^{\mu\nu\lambda\rho} - \nth{90} \; R_{\mu\nu} R^{\mu\nu} +
\nth{144} \;  R^2 + \nth{60} \; \Square \, R \right]  \cr
& \qquad\qquad \qquad\qquad 
+ \; \nth{3} \; C\left(\Scr_\hf\right) \; F^a_{\mu\nu} F_a^{\mu\nu}.\cr}
\eeq
Again $N_\hf$ denotes the total number of Majorana spinors involved.

\topic{Spin One}
Gauge potentials, $A^a_\mu$, are the fields representing spin-one 
particles. Their contribution to the one-loop effective action takes
a form for which the above-described formalism applies
provided that the gauge is chosen appropriately. For 
background-covariant Feynman gauge the appropriate differential
operator for $A^a_\mu$ has the form $-\, \delta^\mu_\nu \, \delta^a_b \;
\Square + \delta^\mu_\nu \, (m_1^2)^{a}_{b} + X^{a\mu}_{b\nu}$, 
with $X^{a\mu}_{b\nu}$
and the traced commutator, $Y_{\mu\nu}$, given in this case by:
\label\Ymnspinone
\eq
\eqalign{
X^{a\mu}_{b\nu} &= - \delta^a_b \; R^\mu_\nu + 2i \, {F^{c\mu}}_{\nu} \, 
\left(\tau_c\right)^a_b \cr
\tr \; \left( Y_{\mu\nu} Y^{\mu\nu} \right) &= - N_1 \; R_{\mu\nu\lambda\rho}
R^{\mu\nu\lambda\rho} + 4 C\left( A \right)  F^a_{\mu\nu} F_a^{\mu\nu}. \cr}
\eeq
Here $\left(\tau_a \right)^b_c = i {c^b}_{ac}$ denotes the gauge generator
in the adjoint representation, and $C(A)$ is the corresponding Dynkin index.
(For $SU(N_c)$ we have $C(A) = N_c$.)

In addition to the vector potentials, each spin-one particle also requires a
Fadeev-Popov-DeWitt ghost, which contributes like a minimally-coupled
complex scalar in the adjoint representation, however it is an anticommuting
field. Combining the vector and ghost contributions gives, for $N_1$ spin-one
particles:
\label\akspinonea
\eq
\tr{}_1\; a_0(x,x) = 2 N_1 , \qquad
\tr{}_1 \; a_1(x,x) = \frac23 \; N_1  \; R , 
\eeq
\label\akspinoneb
\eq
\eqalign{
\tr{}_1\; a_2(x,x) &= N_1 \; \left[ -\; \frac{13}{180} \; R_{\mu\nu\lambda\rho}
R^{\mu\nu\lambda\rho} + \frac{22}{45} \; R_{\mu\nu} R^{\mu\nu} -
\frac{5}{36} \;  R^2 + \nth{10} \; \Square \, R \right]  \cr
& \qquad\qquad \qquad\qquad 
+ \; \frac{11}{6} \; C\left(A\right) \; F^a_{\mu\nu} F_a^{\mu\nu}.\cr}
\eeq
\endtopic

\subsection{Combining Results}

We now proceed to the case of interest, $N=4$ SYM theory in four spacetime
dimensions. This corresponds to the choice $N_0 = 6 N_1$, $N_\hf = 4 N_1$
with $N_1 = d_\ssg$ given by the dimension of the gauge group. We also 
choose all particles to transform in the adjoint representation,
$\Scr_0 = \Scr_\hf = A$, and we take $\xi = - \nth6$, as is appropriate for 
conformally-coupled scalars. For this choice, and combining according 
to $\Tr \; a_k = \tr{}_0\; a_k - \tr{}_\hf \; a_k + \tr{}_1 \; a_k$,
we find:
\label\ztaksuma
\eq
\Tr \; a_0(x,x)  = N_0 - 2 N_\hf + 2 N_1 = 0 
\qquad\qquad \Tr \; a_1(x,x)  = -6 \, 
d_\ssg \left( \xi + \nth6 \right) R = 0,
\eeq
\label\ztaksumb
\eq
\eqalign{
\Tr \; a_2(x,x) &= d_\ssg \left\{ \frac{1}{2} \; R_{\mu\nu} R^{\mu\nu}
+ \left[ 3 \left( \xi + \nth6 \right)^2 - \nth6 \right] R^2 
- \left( \xi + \nth6 \right) \, \Square \, R \right\} \cr
&={d_\ssg\over2} \left(  R_{\mu\nu} R^{\mu\nu}
- \nth3 \;  R^2  \right) . \cr}
\eeq
Note that $\Tr a_2$ takes the form familiar from the conformal
anomaly in which the coefficient of $R_{\mu\nu\lambda\rho}
R^{\mu\nu\lambda\rho}$ vanishes \old.\foot\hawk{We thank
Arkady Tseytlin for pointing out an error in our original version
of this formula.}

\subsection{UV Finiteness on Ultrastatic Spacetimes}

As may be seen from the form of the integrals, $c_k$, of eqs.~\genSigform\
and \ztints, the coefficients $\Tr\; a_k(x,x)$, for $k= 0,1,2$ determine the
ultraviolet divergences of the four-dimensional theory ($n=4$).  
The vanishing of the $k=0$ and $k=1$ terms in eqn.~\ztaksuma\  and the fact
that the expression for the $k=2$ term in eqn.~\ztaksumb\  depends only on the
background curvature (due to the cancellation among different spins of terms
proportional to $F^a_{\mu\nu} F^{\mu\nu}_a$)
 is a check on the calculation. 
It shows that this theory, on flat spacetimes, 
is ultraviolet finite at one loop, as expected. We now show
that this cancellation also occurs for {\it arbitrary} four-dimensional
ultrastatic spacetimes. 

The finite-temperature calculation of the next section is performed
for ultrastatic spacetimes. These admit metrics for which coordinates
may be locally chosen to ensure $ds^2 = d\tau^2 + \gamma_{mn} dx^m dx^n$,
where the spatial metric, $\gamma_{mn}$, is independent of $\tau$. Since
any such metric is effectively three-dimensional, both its squared Weyl
tensor and its Gauss-Bonnet integrand must vanish identically:
\label\vanishids
\eq
\eqalign{
C_{\mu\nu\lambda\rho} C^{\mu\nu\lambda\rho} &= R_{\mu\nu\lambda\rho}
R^{\mu\nu\lambda\rho} - 2 \, R_{\mu\nu} R^{\mu\nu} + \nth3 \; R^2 = 0 \cr
\epsilon^{\mu\nu\lambda\rho} \epsilon^{\alpha\beta\gamma\delta}
R_{\mu\nu\alpha\beta} R_{\lambda\rho\gamma\delta} &= 
- R_{\mu\nu\lambda\rho} R^{\mu\nu\lambda\rho} + 4 \, R_{\mu\nu} R^{\mu\nu} 
- R^2 = 0 .\cr}
\eeq
Eq.~\vanishids\ may be used to eliminate the quantities 
$R_{\mu\nu\lambda\rho} R^{\mu\nu\lambda\rho}$ and 
$R_{\mu\nu} R^{\mu\nu}$ in terms of $R^2$, with the result:
\label\reimricc
\eq
R_{\mu\nu\lambda\rho} R^{\mu\nu\lambda\rho} = 
R_{\mu\nu} R^{\mu\nu} = \nth3 \; R^2. 
\eeq

It is now obvious that eq.~\ztaksumb\ vanishes when
eq.~\reimricc\ are used. Thus we see that $N=4$ SYM is UV finite at
one loop on all four-dimensional ultrastatic spacetimes.
This result also extends to background geometries which have
a product structure, \eg $S^2\times S^2$. Again, since the individual
components of the metric are lower dimensional in such a case, one
finds the vanishing of eq.~\vanishids. However, note that
for more general backgrounds, it is quite possible for $\Tr \; a_2(x,x)$
to be nonzero. For instance, for four-dimensional
maximally symmetric spaces $R_{\mu\nu\lambda\rho}
= \nth{12} \; R \; \left(g_{\mu\lambda} g_{\nu\rho} - g_{\mu\rho}
g_{\nu\lambda} \right)$, and so $4 R_{\mu\nu} R^{\mu\nu} = R^2$,
giving 
\label\maxsymres
\eq
\Tr a_2(x,x) = - \;{d_\ssg\over 24} \; R^2 \ne 0 .
\eeq
In such cases logarithmic UV divergences would appear in the effective action.

\subsection{Nonzero Temperature}

The calculation of the ultraviolet part of the
one-loop effective action for nonzero
temperature is performed along lines very similar to the zero-temperature
result just described\dowker \physrep. We do so by restricting to the
ultrastatic metrics of the
previous section, and periodically identifying the Euclidean time 
$\tau$ with period $\beta = 1/T$. The only change required in the
previous calculation is to find the Heat Kernel which
is periodic --- or, for fermions, antiperiodic --- in time. 

Such a kernel may be obtained from the one used above
at zero temperature through the method of images:
\label\pdcK
\eq
K_\pm(t-t',\bfx,\bfx';s) = \sum_{n=-\infty}^\infty (\pm)^n K(t-t' +n\beta,
\bfx,\bfx';s).
\eeq
where, again, the upper (lower) sign is for bosons (fermions). 
This sum may be evaluated within the small-$s$ expansion 
for the ultrastatic spacetimes of interest here because the time-dependence
of $K_0(\tau - \tau' , \bfx, \bfx ; s)$ is quite simple:
\label\Kzerolim
\eq
K_0(\tau - \tau' , \bfx, \bfx ; s) = K_0(\bfx, \bfx ; s) \;
\exp\left[  - \; {(\tau - \tau')^2 \over 4 s} \right] .
\eeq
Expanding $K_\pm$ for small $s$, as in eq.~\akdefs, and using 
the known form for $K_0$, one finds \dowker:
\label\Ksum
\eq
K_\pm(x,x;s) = K_0(\bfx,\bfx;s) \sum_{n=-\infty}^\infty (\pm)^n
\exp\left( - \; {n^2 \beta^2 \over 4 s} \right).
\eeq
The sums may be performed, giving standard Jacobi $\vartheta$-functions
\wandw:
\label\thetaK
\eq
K_+(x,x;s) = \vartheta_3(\tau) \; K_0(\bfx,\bfx;s), \qquad
K_-(x,x;s) = \vartheta_4(\tau) \; K_0(\bfx,\bfx;s),
\eeq
where $\tau = i\beta^2/(4 \pi s)$. 

Substituting eq.~\thetaK\ into expression \HKrep\
for the effective action, one then finds that the free energy density
is:
\label\genFEform
\eq
F = \mp \; {1 \over 2 \, (4 \pi)^{n/2}} \; 
\sum_{k=0}^\infty C^{(\pm)}_k  \;  \tr \; a_k(x,x) , 
\eeq
where now $C^{(\pm)}_k$ represent the following integrals:
\label\ftints
\eq
\eqalign{
C^{(+)}_k &= \int_0^\infty {ds \over s} \;\; s^{k - n/2} \; \exp(- m^2 s) 
\; \vartheta_3\left({i\beta^2 \over 4 \pi s} \right), \cr
C^{(-)}_k &= \int_0^\infty {ds \over s} \;\; s^{k - n/2} \; \exp(- m^2 s) 
\; \vartheta_4\left({i\beta^2 \over 4 \pi s} \right) .\cr}
\eeq

These integrals diverge for $k=0,1$ and 2 if $n =4$ due to the limit 
of integration at $s \to 0$. This infinity is an ultraviolet 
divergence, and is removed if one focuses
purely on the temperature-dependent part of the problem. This
is accomplished by computing $\Delta^{(\pm)}_k = C^{(\pm)}_k - c_k$, 
which gives the change of the free energy relative to the zero-temperature
result: $\Delta F = F(T) - F(0)$. 

The potential infrared divergence as $s \to \infty$ does not arise so 
long as $m$ is nonzero, so $\Delta^{(\pm)}_k$ may be evaluated 
numerically to any desired accuracy. For our later purposes, however,
it is instructive to examine the high-temperature limit, $m \ll T$. This
limit can be subtle due to the shadow cast by 
incipient infrared divergences which 
can re-emerge in this regime. A well-known example of this occurs
in flat space, since $\Delta^{(+)}_0$ acquires terms such as $m^3 T$ 
when expanded in powers of $m/T$ \WbgFT, \FTeffact. This nonanalytic 
dependence of $\Delta^{(+)}_0$ on $m^2$ as $m^2 \to 0$ thwarts its 
evaluation {\it via} term-by-term integration after expanding its 
integrand in powers of $m^2$, since the successive integrals which are
obtained in this way diverge more and more severely in the infrared.
Such nonanalytic dependence on $m^2$ is characteristic of the 
singularities which arise in the presence of massless particles.

For nontrivial background fields we see that evaluating at
$m = 0$ causes the $\Delta^{(\pm)}_k$ to diverge at
$s \to \infty$ for sufficiently large $k$. This is very similar
to the divergences which are found when a series in 
powers of $m^2$ is attempted in the absence of background fields. 
In this case the divergences are a consequence of a breakdown of the 
derivative expansion due to the nonanalytic dependence on
the background fields which is generated as the curvatures
cut off the infrared singularities which are generated by 
the various massless modes \yuri.

In four dimensions these divergences potentially arise at $m=0$
when $k \ge 2$. We now explicitly display the first of these
by evaluating the terms $k \le 2$ of the Gilkey-De Witt expansion.
One finds a result expressed in terms
of Euler's $\Gamma$-function and Riemann's $\zeta$-function:
\label\ftresult
\eq
\eqalign{
\Delta^{(+)}_k &= \int_0^\infty {ds\over s} \;\; s^{k - n/2} 
\left[ \vartheta_3\left({i\beta^2 \over 4 \pi s} \right) - 1
\right]  
= 2 \left( {4 \over \beta^2} \right)^{{n\over 2} - k} \Gamma\left(
{n\over 2} - k \right) \zeta(n-2k) , \cr 
\Delta^{(-)}_k &= \int_0^\infty {ds\over s} \;\; s^{k - n/2} 
\left[ \vartheta_4 \left({i\beta^2 \over 4 \pi s} \right) - 1
\right]  
= \left( 2^{1-n+2k} -1 \right) \; \Delta^{(+)}_k.  \cr}
\eeq

Combining all expressions, and expanding about $n=4$, 
gives the following result for the 
free energy density of a general theory:
\label\FEresult
\eq
\eqalign{
\Delta F  &= - \; {1 \over 2 \, (4 \pi)^{n/2}} \; 
\sum_{k=0}^\infty \Delta^{(+)}_k \; \left[ \tr{}_\ssb \; a_k(x,x) 
+ \left(1- 2^{1-n+2k}  \right) \; \tr{}_\ssf \; a_k(x,x) \right],  \cr
&=  - \; {\pi^2 T^4 \over 90} \left[ \tr{}_\ssb \; a_0 + \frac78 \; 
\tr{}_\ssf \; a_0 \right] - \; {T^2 \over 24} \; 
\left[ \tr{}_\ssb \; a_1 + \frac12 \; \tr{}_\ssf \; a_1 \right] \cr
& \qquad- \; {1 \over 32 \pi^2} \left[ {2 \over 4-n}
- 2 \ln \left( {T \over T_0} \right) + \cdots \right] 
 \left[ \tr{}_\ssb \; a_2 - \tr{}_\ssf \; a_2 \right] +
 {3 \ln 2 \over 32 \pi^2} \; \tr{}_\ssf \; a_2 + \cdots . \cr }
\eeq
Here $\tr{}_\ssb \; a_k$ and $\tr{}_\ssf \; a_k$ respectively denote
the trace over the bosons and fermions of the model. The ellipses
within the square brackets represent terms which neither diverge,
nor involve the logarithm of $T$. By contrast, the ellipses at the end of the
equation represent terms which involve higher than two powers of curvature
and gauge field strengths. 

The infrared divergence arises in eq.~\FEresult\
as the pole as $n \to 4$ in the
$k=2$ term of the series. Notice that the condition for the cancelling of this
particular divergence is $ \tr{}_\ssb \; a_2 = \tr{}_\ssf \; a_2$,
which is precisely the condition for the cancelling
of the zero temperature logarithmic ultraviolet divergence.
As a consequence this particular divergence does not arise in the 
present example of $N=4$ SYM, although the same is not expected to be
true for the infrared divergences arising for higher $k$.

Evaluating this expression using the $a_k$'s of $N=4$ SYM, and
using the simplifications, eq.~\reimricc, which follow from
the restriction to an ultrastatic spacetime, finally
gives:
\label\finalsymform
\eq
\Delta F =  - \; {\pi^2 d_\ssg T^4 \over 6}
- \; {d_\ssg T^2 R \over 24} + {d_\ssg  \ln 2 \over 160 \pi^2} \; \Square \, R
+ {C(A) \ln 2 \over 8 \pi^2 } \; F^a_{\mu\nu} F_a^{\mu\nu} + \cdots  
\eeq

In the next section we compute the free energy density in the strong-coupling
limit using the AdS/CFT correspondence \juanA. We do so for the special case
of vanishing  background gauge fields, and for 
the specific case where the spatial three-geometry has constant curvature
The Ricci scalar for such a space may be written
as $R = - 6\kappa/\ell^2$,
where $\ell$ is the radius of curvature, and $\kappa=+1,0,-1$ for a three-sphere,
flat space and a hyperbolic three-plane. Further
for comparison purposes, we will be interested in the gauge group
for $SU(N_c)$ in the large-$N_c$ limit, so $d_\ssg = N_c^2 -1
\approx N_c^2$. 
For these choices, our result in eq.~\finalsymform\ yields:
\label\compform
\eq
\eqalign{
\Delta F &=  N_c^2  \left[ - \; {\pi^2  T^4 \over 6}
+ \; { \kappa T^2 \over 4 \ell^2}  + O\left({1\over\ell^6T^2}\right) \right]\cr
&=-\; {\pi^2 N_c^2T^4 \over 6}\left[1-\,{3 \kappa\over 2\pi^2 \ell^2T^2}+
O\left({1\over\ell^6T^6}\right)\right].\cr}
\eeq

\section{Strong-Coupling Calculation}

Using the AdS/CFT correspondence, the dual supergravity
description of the $N=4$ SYM theory
at finite temperature is an asymptotically anti-de Sitter
black hole \juanA\therm. An appropriate class of metrics describing
Euclideanized black holes is \birmbh
\label\blackmet
\eq
ds^2=\left({r^2\over\ell^2}+\kappa -{r_0^2\over r^2}\right)d\tau^2
+{dr^2\over{r^2\over\ell^2}+\kappa -{r_0^2\over r^2}}+{r^2\over\ell^2}
\; d\Sigma_3(\kappa)
\eeq
where $\kappa=+1,0,-1$ and $d\Sigma_3(k)$ is the 
line element on a three-dimensional
manifold with constant curvature $\kappa/\ell^2$. An explicit representation of
the latter may be chosen as
\label\three
\eq
d\Sigma_3(\kappa)=\ell^2\left[(1-\kappa\rho^2)dz^2 +{d\rho^2\over1-
\kappa\rho^2}+
\rho^2d\phi^2\right]
\eeq
but the precise form of the three-dimensional metric will not be needed
in the following. All of these five-dimensional metrics \blackmet\ satisfy
the equation of motion,\foot\convchange{Note
that there is a change of conventions for the curvatures between this
section and the previous one. Those of this section follow those of
ref.~\mtw, since these are conventional in the gravity literature.
Although the metric signature remains $\ss -+++$, we must
replace $\ss R_{\mu\nu\lambda\rho}
\to -  R_{\mu\nu\lambda\rho}$ in all formulae.}
$R_{\mu\nu}=-(4/\ell^2)g_{\mu\nu}$, which arises from the action
\label\einact
\eq
I=-{1\over16\pi G_5}\int d^5x \sqrt{g}(R+12/\ell^2)
-{1\over8\pi G_5}\int d^4x \sqrt{h}K\ .
\eeq
To avoid a conical singularity at 
\label\orig
\eq
r_+={1\over2}\left(\sqrt{\kappa^2\ell^2+4r_0^2\ell^2}-\kappa\ell\right)\ ,
\eeq
where $g_{\tau\tau}$ vanishes, one must select the period of $\tau$
to be
\label\tperiod
\eq
{1\over T}=\beta={2\pi\ell^2r_+\over 2r_+^2+\kappa\ell^2}\ .
\eeq
Now, the Euclidean action evaluated for the classical black hole metric
is interpreted as giving the leading contribution
to the free energy. In the context of AdS black holes, such calculations
were first carried out in ref.~\don. As typically arises in these calculations,
however, this action diverges and so to produce a finite result,
we subtract off the contribution of a reference metric \garyB.
This step is analogous to subtracting off the zero-temperature
free energy in the field theory calculation --- see section 3.5.
In this case, the reference geometry is simply anti-de Sitter
space which is produced by setting $r_0=0$ in eq.~\blackmet. Taking
care to match the asymptotic geometries correctly, one finds
\label\daction
\eq
\Delta I=V_3 \beta\,\Delta F=
V_3 \beta\, {r_+^2(\ell^2\kappa-r_+^2)\over16\pi G_5\ell^5}\ .
\eeq
Note that in defining $\Delta I=V_3 \beta\Delta F$, the relevant volume
is essentially the coordinate volume of $\tau$ and $d\Sigma_3(\kappa)$. It
is not the proper volume of a surface of constant radius at large $r$
in eq.~\blackmet, which of course diverges as $r\rightarrow\infty$.
 
Now the duality prescription \juanA\ identifies
$G_5\ell^5=8\pi^3g^2\alpha'{}^4$
and $\ell^4=4\pi g N_c\alpha'{}^2$, where $g$ and $\alpha'$ are the string
coupling constant and the inverse string tension, respectively.
Combining these formula then yields  the free energy density
at strong coupling for $SU(N_c)$ SYM in the large-$N_c$ limit:
\label\strcpresult
\eq
\Delta F= - \; {\pi^2 N_c^2 T^4 \over 8} \; \Scf\left({\kappa \beta^2 \over
\ell^2} \right),
\eeq
where the function $\Scf(x)$ is given by:
\label\exactfn
\eq
\eqalign{
\Scf(x) &= \nth{16} \; \left[ 1 + \left(1 - {2 x\over \pi^2} \right)^\hf
\right]^2 \left\{ \left[ 1 + \left( 1 - { 2x\over \pi^2} \right)^\hf
\right]^2 - {4x \over \pi^2} \right\} \cr
&\approx 1 - 3 \; \left( {x \over \pi^2} \right)
+ {3\over2} \; \left({x \over \pi^2} \right)^2 
+ \nth4 \;  \left({x \over \pi^2} \right)^3   + \cdots \cr}
\eeq
where the final expansion applies for $x<<1$. This expansion corresponds
precisely to the high temperature limit in eq.~\strcpresult, which then
becomes
\label\strcpresulta
\eq
\eqalign{
\Delta F
& = N_c^2 \left[ - \; {\pi^2   T^4 \over 8} + { 3\kappa T^2 \over 8 \ell^2} 
  -{3\kappa^2\over16\pi^2\ell^4}+O\left({1\over\ell^6T^2}\right)\right] \cr
& = -{\pi^2\over8}N_c^2T^4\left[1-{3\over\pi^2}\,{\kappa\over\ell^2T^2}+
{3\over2\pi^4}\,{\kappa^2\over\ell^4T^4}+O\left({1\over\ell^6T^6}\right)
\right].\cr}
\eeq
The latter is now in a form which is readily compared to the weak coupling
result \compform.

For the $\kappa=0$ case, of course, the leading term is the full answer up
to corrections in the effective coupling. For the $\kappa=+1$ and --1 cases,
similar calculations yielding the curvature corrections appear in
refs.~\garyA\ and \roberto, respectively.

Notice that for $\kappa=0$ or --1, $\Delta I$ in eq.~\daction\
is always negative, and so the black hole solution \blackmet\ always
provides the dominant saddle-point in the supergravity path integral
for any temperature. On the other hand, for $\kappa=+1$, $\Delta I$
becomes positive for $r_+<\ell$, which corresponds to
$\beta>2\pi\ell/3$. Hence in this low temperature range, AdS space
itself is the dominant saddle-point. This change at $\beta=2\pi\ell/3$
is believed to be associated with a deconfinement 
phase transition in the SYM theory
\therm.\foot\branch{Note that below this phase transition at
$\ss \beta=\pi\ell/\sqrt{2}$, there is a branch point in
$\ss \Scf(\beta^2/\ell^2)$.
This corresponds to the minimum temperature for which a black hole
solution with $\ss \kappa=+1$ exists.}
For comparision to the weak-coupling calculation, we are interested
in the high temperature regime where the black hole is the dominant
AdS saddle-point for all three choices, $\kappa=0,\pm 1$.

\section{Discussion}

In comparing the free energy densities calculated at one loop
\compform\  and strong coupling \strcpresult, we see that these 
expressions, while not identical, yield two expansions 
which look very similar. Both give a free energy having 
a high-temperature limit which has the form of a 
Taylor expansion in powers of $\beta^2/\ell^2$:
\label\hammer
\eq
\Delta F=-\; {\pi^2 N_c^2T^4\over6}\sum_{n=0}^\infty b_n(\lambda) \left(
{2 \kappa\over\pi^2\ell^2T^2}\right)^n
\eeq
where the coefficients $b_n(\lambda)$ are functions of
the effective 't Hooft coupling, $\lambda=g_{\rm YM}^2N_c=
\ell^4/(2\alpha'^2)$. 

The significance of this observation for the strong-coupling expansion
is in what eq.~\hammer\ does {\it not} contain. In particular,
the weak-coupling expansion had the potential to involve both
fractional powers of $\beta^2/\ell^2$ (similar to the $m^3T$ term
in $\Delta^{(+)}_0$), as well as terms proportional to 
$\log \beta$ (such as when $\Tr a_2 \ne 0$). Such terms
reflect the infrared divergences which are associated with 
the massless bosons of the perturbative spectrum. They
do not arise at one loop in the weak-coupling calculation
due to the cancellations in $\Tr a_2$, which reflect the finiteness
and conformal invariance of $N=4$ SYM theory in ultrastatic
background geometries. This same cancellation does not hold in 
general for all terms in the weak-coupling limit. For instance, 
even on flat space the massless gauge bosons generate infrared 
divergences to perturbative calculations of thermal quantities like
gluon damping rates \gluondamping, once one proceeds 
beyond one loop. The same might also be expected to follow for the 
$k\ge 3$ terms of the derivative expansion at one loop.

Unlike the weak-coupling result, the strong-coupling expression,
eq.~\exactfn\ is not simply given as an asymptotic form for
large temperatures, but is explicitly given in terms of the function
$\Scf(\kappa\beta^2/\ell^2)$ for the geometries of interest.
Explicit examination of $\Scf(x)$ clearly shows it to be analytic 
in its argument near $x=0$, indicating the absence of singularities 
in the high-temperature regime. But the existence of exact 
conformal invariance of the underlying model ensures the absence 
of a gap in the particle spectrum. In four dimensions this
follows directly if the ground state respects the
conformal invariance, or from the existence of a Goldstone `dilaton' 
mode if the conformal symmetry is spontaneously broken. Infrared
divergences would be expected to be weakened or absent if the only
massless states were Goldstone modes \GBIR, since these decouple in the
long-wavelength limit \GBreview. The absence of infrared singularities 
in the large temperature limit of the strong-coupling calculation
therefore indicates either the existence of only Goldstone massless
modes, or the persistence into the large-$N_c$ limit of the one-loop 
cancellations of infrared divergences which were seen at weak coupling
for non-Goldstone massless modes. 

Notice there is no singularity predicted by eq.~\exactfn\ for any 
other temperatures, in spite of the branch point which apparently
arises at $x = \pi^2/2$. This branch point has no direct implications
because: $(i)$ $x = - \beta^2/\ell^2 \le 0$ if $\kappa = -1$;
$(ii)$ the function $\Scf(x)$ collapses to a constant 
(and hence the branch point vanishes) if $\kappa = 0$;
and $(iii)$ the branch point lies outside the high temperature
domain of validity 
($\beta < 2 \pi \ell/3$) of the strong-coupling calculation if
$\kappa = +1$.

The behaviour we find is consistent with the behaviour
which was previously argued for $N=4$ SYM by Witten \therm. 
It would be of interest to extend other techniques which have proven
powerful in flat space, such as those of ref.~\Dorey, to analyse the 
curved spaces considered here in more detail. In particular, 
so far as they go our results are consistent with a QCD-like 
picture in which no phase transition occurs for any finite 
coupling, $\lambda_c$. 

Although our calculation cannot reveal the functional form
for $b_n(\lambda)$ away from the limits $\lambda \to 0$
and $\lambda \to \infty$, it does differ in some ways from
similar calculations of the free energy density
in the Higgs phase of the SYM theory \arkady. In that case, 
the results suggest that the strong-to-weak
coupling interpolation is achieved with a single overall 
multiplicative function.
While one might have hoped for a similar simple form in the present
calculation, it is clear from
our expressions that this is not the case. Rather,
the weak coupling form
\compform\ determines
\label\zero
\eq
b_0(0)=1\ ,\qquad\quad b_1(0)={3\over4}\ ,\qquad\quad
b_2(0)=0\ ,
\eeq
while the strong coupling expansion \strcpresult\ fixes
\label\biglam
\eq
b_0(\lambda\rightarrow\infty)\rightarrow {3\over4}\ ,\qquad
b_1(\lambda\rightarrow\infty)\rightarrow {3\over2}\ ,\qquad
b_2(\lambda\rightarrow\infty)\rightarrow {3\over8}\ .
\eeq

Following the arguments presented in ref.~\garyA,
the free energy densities calculated
using Euclidean techniques above also have an interpretation as Casimir
energy densities. The latter arises if the Wick rotation back to a Minkowski
signature manifold is made on an appropriate `time' coordinate
in the constant curvature three-manifold. Then
eqs.~\compform\ and \strcpresult\ yield the Casimir energy density for SYM
on $dS_3\times S^1$, $R^3\times S^1$ and $AdS_3\times S^1$ for $\kappa=+1,$
0 and $-1$, respectively. The negative Casimir energy is generated by
antiperiodic boundary conditions on the $S^1$ factor for the fermions.

Finally, in closing, we remark that the calculations in section 3 show
that in a general curved space, $N=4$ super-Yang-Mills need not be
finite. We observed though that for background geometries with
a product structure into two (or more) lower dimensional components,
UV finiteness is maintained due to a remarkable cancellation of 
terms in $\Tr a_2$, which is given in eq.~\ztaksumb.
Indeed, in the finite temperature calculations, finiteness arose because of
the ultrastatic form of the background metric, which gives a product
structure of the form $S^1\times M_3$.
Of course, it is not difficult to find backgrounds where no
such cancellation occurs. For example, consider the simple
case of $S^4$, which
is maximally symmetric with $R = 12/\ell^2$ where $\ell$ 
is the radius of curvature. In this case eq.~\maxsymres\ implies
the nonzero result
\label\nocancel
\eq
\Tr \; a_2= - \; 6\,{N_c^2\over\ell^4} ,
\eeq
and hence a logarithmic UV divergence appears in the SYM effective action.

It is not difficult to choose coordinates on five-dimensional AdS space
such that the `boundary' manifold (\ie the asymptotic regulator surfaces)
is $S^4$, \ie
\label\newmeter
\eq
ds^2={dr^2\over1+{r^2\over\ell^2}} +r^2 d\Omega_4
\eeq
where $d\Omega_4$ is the standard metric on a unit four-sphere.
Now the logarithmic UV divergence appearing at one-loop
in the SYM effective action would at first sight seem to present
a problem for the proposed AdS/CFT correspondence, but in
fact, it provides a remarkable consistency check \boundtermb.
The AdS/CFT correspondence provides a new set of intrinsic surface terms
\boundterms,\boundtermb\ (for related work,
see \boundman) for the gravitational action \einact. Remarkably, these 
new terms render the gravitational action finite except precisely in the
case where $\Tr  a_2\not=0$ for the boundary surface. In the latter case,
the volume integral produces a logarithmic divergence instead. For
the $S^4$ boundary, one can precisely match this divergence with that
appearing in the dual field theory calculation \boundtermb. Hence, in
keeping with the UV/IR relation \uvir\  of the AdS/CFT  correspondence, 
one again
finds that an ultraviolet divergence in the field theory is matched by an
infrared divergence in the supergravity.

\bigskip

\centerline{\bf Acknowledgments}

\bigskip

This research was partially funded by N.S.E.R.C.\ of Canada and les  
Fonds F.C.A.R.\ du Qu\'ebec. We would like to thank Yuri Gusev, Marcia
Knutt and Guy Moore for useful conversations.

\bigskip

\listrefs

\bye

%% file: macros.tex

\font\titlefont = cmr10 scaled\magstep 4
 2
\font\sectionfont = cmr10
\font\littlefont = cmr5 
\font\eightrm = cmr8 

\def\ss{\scriptstyle} 
\def\sss{\scriptscriptstyle} 

\newcount\tcflag
\tcflag = 0  

\ifnum\tcflag = 0 \magnification = 1200 \fi  

\global\baselineskip = 1.2\baselineskip 
\global\parskip = 4pt plus 0.3pt 
\global\abovedisplayskip = 18pt plus3pt minus9pt
\global\belowdisplayskip = 18pt plus3pt minus9pt
\global\abovedisplayshortskip = 6pt plus3pt
\global\belowdisplayshortskip = 6pt plus3pt

\def\barsoff{\overfullrule=0pt}


\def\endignore{}
\def\ignore #1\endignore{} 

\newcount\dflag
\dflag = 0


\def\monthname{\ifcase\month 
\or January \or February \or March \or April \or May \or June%
\or July \or August \or September \or October \or November %
\or December 
\fi}

\newcount\dummy
\newcount\minute  
\newcount\hour
\newcount\localtime
\newcount\localday
\localtime = \time
\localday = \day

\def\advanceclock#1#2{ 
\dummy = #1
\multiply\dummy by 60
\advance\dummy by #2
\advance\localtime by \dummy
\ifnum\localtime > 1440 
\advance\localtime by -1440
\advance\localday by 1
\fi}

\def\settime{{\dummy = \localtime %
\divide\dummy by 60%
\hour = \dummy 
\minute = \localtime%
\multiply\dummy by 60%
\advance\minute by -\dummy 
\ifnum\minute < 10 
\xdef\spacer{0} 
\else \xdef\spacer{} 
\fi %
\ifnum\hour < 12 
\xdef\ampm{a.m.} 
\else 
\xdef\ampm{p.m.} 
\advance\hour by -12 %
\fi %
\ifnum\hour = 0 \hour = 12 \fi 
\xdef\timestring{\number\hour : \spacer \number\minute%
\thinspace \ampm}}}



\def\endtitle{}
\def\title#1\endtitle{\vskip.5in\titlefont
\global\baselineskip = 2\baselineskip 
#1\vskip.4in
\baselineskip = 0.5\baselineskip\rm}
 
\def\endauthors{}
\def\authors#1\endauthors{#1}

\def\endabstract{}
\def\abstract#1\endabstract{\vskip .3in%
\centerline{\sectionfont\bf Abstract}%
\vskip .1in
\noindent#1}

\def\nopageonenumber{\footline={\ifnum\pageno<2\hfil\else
\hss\tenrm\folio\hss\fi}}  

\newcount\nsection 
\newcount\nsubsection 

\def\section#1{\global\advance\nsection by 1
\nsubsection=0
\bigskip\noindent\centerline{\sectionfont \bf \number\nsection.\ #1}
\bigskip\rm\nobreak}

\def\subsection#1{\global\advance\nsubsection by 1
\bigskip\noindent\sectionfont \sl \number\nsection.\number\nsubsection)\
#1\bigskip\rm\nobreak}

\def\topic #1{{\medskip\noindent $\bullet$ \it #1:}} 
\def\endtopic{\medskip}

\def\appendix#1#2{\bigskip\noindent%
\centerline{\sectionfont \bf Appendix #1.\ #2} 
\bigskip\rm\nobreak} 


\newcount\nref 
\global\nref = 1 

\def\therefs{} 


\def\ref#1#2{\xdef #1{[\number\nref]} 
\ifnum\nref = 1\global\xdef\therefs{\item{[\number\nref]} #2\ } 
\else
\global\xdef\oldrefs{\therefs}
\global\xdef\therefs{\oldrefs\vskip.1in\item{[\number\nref]} #2\ }%
\fi%
\global\advance\nref by 1
}

\def\listrefs{\vfill\eject\section{References}\therefs}


\newcount\nfoot 
\global\nfoot = 1 

\def\foot#1#2{\xdef #1{(\number\nfoot)} 
\hskip -0.2cm ${}^{\number\nfoot}$ 
\footnote{}{\vbox{\baselineskip=10pt
\eightrm \hskip -1cm ${}^{\number\nfoot}$ #2}}
\global\advance\nfoot by 1
}


\newcount\nfig 
\global\nfig = 1
\def\thefigs{} 

\def\figure#1#2{\xdef #1{(\number\nfig)}
\ifnum\nfig = 1\global\xdef\thefigs{\item{(\number\nfig)} #2\ }
\else
\global\xdef\oldfigs{\thefigs}
\global\xdef\thefigs{\oldfigs\vskip.1in\item{(\number\nfig)} #2\ }%
\fi%
\global\advance\nfig by 1 } 

\def\fig#1{\xdef #1{(\number\nfig)}
\global\advance\nfig by 1 } 


\newcount\ntab
\global\ntab = 1

\def\table#1{\xdef #1{\number\ntab}
\global\advance\ntab by 1 } 


\newcount\cflag
\newcount\nequation
\global\nequation = 1
\def\eqlabel{(1)}

\def\nexteqno{\ifnum\cflag = 0
\global\advance\nequation by 1
\fi
\global\cflag = 0
\xdef\eqlabel{(\number\nequation)}}

\def\lasteqno{\global\advance\nequation by -1
\xdef\eqlabel{(\number\nequation)}}

\def\label#1{\xdef #1{(\number\nequation)}
\ifnum\dflag = 1
{\escapechar = -1
\xdef\draftname{\littlefont\string#1}}
\fi}

\def\clabel#1#2{\xdef\eqlabel{(\number\nequation #2)}
\global\cflag = 1
\xdef #1{\eqlabel}
\ifnum\dflag = 1
{\escapechar = -1
\xdef\draftname{\string#1}}
\fi}

\def\cclabel#1#2{\xdef\eqlabel{#2)}
\global\cflag = 1
\xdef #1{\eqlabel}
\ifnum\dflag = 1
{\escapechar = -1
\xdef\draftname{\string#1}}
\fi}


\def\eeq{}

\def\eqnn #1\eeq{$$ #1 $$}

\def\eq #1\eeq{
\ifnum\dflag = 0
{\xdef\draftname{\ }}
\fi 
$$ #1
\eqno{\eqlabel \rlap{\ \draftname}} $$
\nexteqno}







\def\eqa #1\eeq{
\ifnum\dflag = 0
{\xdef\draftname{\ }}
\fi 
$$ \eqalignno{ #1 } $$
\global\cflag = 0}


\def\ie{{\it i.e.\/}}
\def\eg{{\it e.g.\/}}



\global\nulldelimiterspace = 0pt



\def\frac#1#2{{{#1} \over {#2}}\,}  
\def\hf{{1\over 2}}
\def\nth#1{{1\over #1}}


\def\Square{{\vbox {\hrule height 0.6pt\hbox{\vrule width 0.6pt\hskip 3pt
        \vbox{\vskip 6pt}\hskip 3pt \vrule width 0.6pt}\hrule height 0.6pt}}}
\def\Asl{\hbox{/\kern-.7500em\it A}} 
\def\Dsl{\hbox{/\kern-.6700em\it D}} 
\def\dsl{\hbox{/\kern-.5300em$\partial$}}
\def\pxpsl{\hbox{/\kern-.5600em$p$}}
\def\sslsh{\hbox{/\kern-.5300em$s$}}
\def\epssl{\hbox{/\kern-.5100em$\epsilon$}}
\def\delsl{\hbox{/\kern-.6300em$\nabla$}}
\def\lxpsl{\hbox{/\kern-.4300em$l$}}
\def\elxpsl{\hbox{/\kern-.4500em$\ell$}}
\def\kxpsl{\hbox{/\kern-.5100em$k$}}
\def\qxpsl{\hbox{/\kern-.5000em$q$}}
\def\sla#1{\raise.15ex\hbox{$/$}\kern-.57em #1}



\def\roughly#1{\mathrel{\raise.3ex\hbox{$#1$\kern-.75em\lower1ex\hbox{$\sim$}}}}



\def\bfx{{\bf x}}



\def\Scf{{\cal F}}

\def\Scr{{\cal R}}

\def\Scw{{\cal W}}


\def\ssb{{\sss B}}

\def\ssf{{\sss F}}
\def\ssg{{\sss G}}

\def\ssl{{\sss L}}

\def\ssr{{\sss R}}


\def\pmb#1{\setbox0=\hbox{#1}%
\kern-.025em\copy0\kern-\wd0
\kern.05em\copy0\kern-\wd0
\kern-.025em\raise.0433em\box0}   


\font\jlgtenbrm=cmbx10
\font\jlgtenbit=cmmib10
\font\jlgtenbsy=cmbsy10
\font\jlgsevenbrm=cmbx10 at 7pt
\font\jlgsevenbsy=cmbsy10 at 7pt
\font\jlgsevenbit=cmmib10 at 7pt
\font\jlgfivebrm=cmbx10 at 5pt
\font\jlgfivebsy=cmbsy10 at 5pt
\font\jlgfivebit=cmmib10 at 5pt
\newfam\jlgbrm

\textfont\jlgbrm=\jlgtenbrm
\scriptfont\jlgbrm=\jlgsevenbrm
\scriptscriptfont\jlgbrm=\jlgfivebrm
\newfam\jlgbit

\textfont\jlgbit=\jlgtenbit
\scriptfont\jlgbit=\jlgsevenbit
\scriptscriptfont\jlgbit=\jlgfivebit
\newfam\jlgbsy

\textfont\jlgbsy=\jlgtenbsy
\scriptfont\jlgbsy=\jlgsevenbsy
\scriptscriptfont\jlgbsy=\jlgfivebsy
\newcount\jlgcode
\newcount\jlgfam
\newcount\jlgchar
\newcount\jlgtmp
\def\bolded#1{
        \jlgcode\the#1 \divide\jlgcode by 4096
        \jlgtmp\the\jlgcode \multiply\jlgtmp by 4096
        \jlgfam\the#1 \advance\jlgfam by -\the\jlgtmp
        \divide\jlgfam by 256
        \jlgtmp\the\jlgcode \multiply\jlgtmp by 16
	\advance\jlgtmp by \the\jlgfam
	\multiply\jlgtmp by 256
        \jlgchar\the#1 \advance\jlgchar by -\the\jlgtmp
        \advance\jlgfam by \the\jlgbrm
        \jlgtmp\the\jlgcode
        \multiply\jlgtmp by 16
        \advance\jlgtmp by \the\jlgfam
        \multiply\jlgtmp by 256
        \advance\jlgtmp by \the\jlgchar
        \mathchar\the\jlgtmp
}


\def\tr{\mathop{\rm tr}}
\def\Tr{\mathop{\rm Tr}}

\def\Log{\mathop{\rm Log}}




